\documentclass[11pt,a4paper]{article}
\pdfoutput=1 

\usepackage{jheppub} 

\usepackage[T1]{fontenc} 

\title{\boldmath Mass spectra and decay properties of $D$ Meson  in a relativistic Dirac formalism}

\author[a,b]{Manan Shah,}
\author[b]{Bhavin Patel,}
\author[a]{P C Vinodkumar}

\affiliation[a]{Department of Physics, Sardar Patel University,\\ Vallabh Vidyanagar - 388 120, INDIA}
\affiliation[b]{P. D. Patel Institute of Applied Sciences, CHARUSAT,\\ Changa -  388 421, INDIA}

\emailAdd{mnshah09@gmail.com}
\emailAdd{azadpatel2003@gmail.com}
\emailAdd{p.c.vinodkumar@gmail.com}

\abstract{The mass spectra of $D$ meson states are calculated in the framework of a relativistic independent quark model. For the present study, we have used the martin like potential for the quark confinement.  Our predicted states in S-wave, $2\ ^3S_1$ (2605.86 MeV) and $2\ ^1S_0$ (2521.72 MeV) are in very good agreement with experimental result of $2608\pm{2.4}\pm{2.5}$ MeV and $2539.4\pm{4.5}\pm{6.8}$ MeV respectively reported by BABAR Collaboration. The calculated P-wave $D$ meson states, $1^3P_2$ (2468.22 MeV), $1^3P_1$ (2404.94 MeV), $1^3P_0$ (2315.24 MeV) and $1^1P_1$ (2367.94 MeV) are in close agreement with experimental average (Particle Data Group) values of $2462.6 \pm 0.7 $ MeV, $2427 \pm 26 \pm 25$ MeV, $2318 \pm 29 $ MeV and $2421.3 \pm 0.6 $ MeV respectively. The pseudoscalar decay constant ($f_P$= 202.57 MeV) of $D$ meson obtained using this relativistic formalism  is in very good agreement with the experiment as well as with the lattice and other available theoretical predictions. The Cabibbo favoured hadronic decay branching ratios, BR$(D^0\rightarrow K^- \pi^+)$ as $3.835 \%$ and BR $(D^0\rightarrow K^+ \pi^-)$ as $1.069 \times 10^{-4} $ are also in very good agreement with the respective experimental values of $ 3.91 \pm 0.08\%$ and $(1.48\pm 0.07) \times 10^{-4}$ reported by CLEO Collaboration. Our predicted results in leptonic decay widths of $D$ meson are also in better accord with experiment as well as other theoretical results. The mixing parameters of $D^0 - \bar{D}^0$ oscillation, $x_q$ (5.14 $\times 10^{-3}$), $y_q$ (6.02 $\times 10^{-3}$) and $R_M$ (3.13 $\times 10^{-5}$) are in very good agreement with BaBar and Belle Collaboration results.}

\begin{document}
\maketitle
\flushbottom

\section{Introduction} \label{intro}
Very recently, experiments at LHCb \cite{Yuan2013} have reported large number of $D_J$ resonances in the mass range of $2.0 GeV /c^2 $ to $ 4.0 GeV/ c^2$ of which many of them belong to natural excited states of $D$ meson while quite a number of them belong to unnatural states \cite{Yuan2013}. It is important and necessary to exhaust the possible conventional description of  $q\bar{Q}$ excitations \cite{PDG2012} before resorting to more exotic interpretations \cite{Wang2008,Vijande2009}. Further theoretical efforts are still required in order to explain satisfactorily the recent experimental data concerning these open-charm states.\\

 Apart from the challenges posed by the exotics, there are also many states which are admixtures of their nearby natural states. For example, the discoveries of new resonances of $D$ states such as $D$(2550) \cite{Zhi2010}, $D$(2610) \cite{Zhi2010}, $D$(2640) \cite{Abreu1998}, $D$(2760) \cite{Zhi2010} etc., have further generated considerable interest towards the spectroscopy of this open charm mesons. Study of $D$ meson carry special interest as it is a hadron with two open flavours ($c, \bar u$ or $\bar d  $) which restricts its decay via strong interactions. These resonance states thus provide us a clean laboratory to study electromagnetic and weak interactions. The masses of low-lying $1S$ and $1P_J$ states of $D$ mesons are recorded both experimentally \cite{PDG2012} and theoretically \cite{Godfrey85,Godfrey,Pierro,Falk,Bhavin2010,Naynesh2013}. Though lattice QCD and QCD sum rule are quite successful, but their predictions for the excited states of the open flavor mesons in the heavy sector are very few. However recent experimental data on excited $D-$ states are partially inconclusive and require more detailed analysis involving their decay properties. The understanding of the weak transition form factors of heavy mesons is important for a proper extraction of the quark mixing parameters, for the analysis of non-leptonic decays and CP violating effects. QCD sum rule (QSR) \cite{Gelhausen2014,Gelhausen2013,Lucha2011,Hayashigaki2004,Lozea2007} is non-perturbative approach to evaluate hadron properties by using the correlator of the quark currents over the physical vacuum and it is implemented with the operator product expansion (OPE). Lattice QCD (LQCD) \cite{Moir2013,Moir2012,Dimopoulos2013} is also non-perturbative approach to use a discrete set of spacetime points (lattice) to reduce the analytically intractable path integrals of the continuum theory to a very difficult numerical computation. QCD sum rules are suitable for describing the low $q^2$ region of the form factors; lattice QCD gives good predictions for high $q^2$. As a result these methods do not provide for a full picture of the form factors and more significant, for the relations between the various decay channels. Potential models provide such relations and give the form factors in the full $q^2$-range.  \\

Thus any attempts towards the understanding of these newly observed states become very important for our understanding of the light quark/antiquark dynamics within $q\bar{Q}$/$Q\bar{q}$ bound states. So, a successful theoretical model aims to provide important information about the quark-antiquark interactions and the behavior of QCD within the doubly open flavour hadronic system. Though there exist many theoretical models \cite{Godfrey85,Godfrey,Pierro} to study the hadron properties based on its quark structure,  the predictions for low-lying states are off by $60-90$ MeV with respect to the respective experimental values. Moreover the issue related to the hyperfine and fine structure splitting of the mesonic states; their intricate dependence with the constituent quark masses and the running strong coupling constant are still unresolved. Though the validity of nonrelativistic models is very well established and significantly successful for the description of heavy quarkonia, disparities exist in the the description of meson containing light flavour quarks or antiquarks. \\

For any successful attempt to understand these states not only be able to satisfactorily predict the mass spectra but also be able to predict their decay properties. For better predictions of the decay widths, many models have incorporated additional contributions such as radiative and higher order QCD corrections \cite{Bhavin2010, Rai2008,Ebert03,Lansberg,Kim}. Thus, in this paper we make an attempt to study properties like mass spectrum, decay constants and other decay properties of the $D$ meson based on a relativistic Dirac formalism. We investigate the heavy-light mass spectra of $D$ meson in this framework with Martin like confinement potential as in the case of $D_s$ mesons studied recently \cite{Manan2014}. \\

Along with the mass spectra, the pseudoscalar decay constants of the heavy-light mesons have also been estimated in the context of many QCD-motivated approximations. The predictions of such methods spread over a wide range of values \cite{Wang,Cvetic}. It is important thus to have reliable estimate of the decay constant as it is an important parameter in many weak processes such as quark mixing, CP violation, etc. The leptonic decay of charged meson is another important annihilation channel through the exchange of virtual $W$ boson. Though this annihilation process is rare, but they have clear experimental signatures due to the presence of highly energetic leptons in the final state. The leptonic decays of mesons entails an appropriate representation of the initial state of the decaying vector mesons in terms of the constituent quark and antiquark with their respective momenta and spin. The bound constituent quark and antiquark inside the meson are in definite energy states having no definite momenta. However one can find the momentum distribution amplitude for the constituent quark and antiquark inside the meson just before their annihilation to a lepton pair. Thus, it is appropriate to compute the leptonic branching ratio and compare our result with the experimental values as well as with the predictions based on other models.

\section{Theoretical Framework}
 The quark confining interaction of meson is considered to be produced by the non-perturbative multigluon mechanism and this mechanism is unfeasible to estimate theoretically from first principles of QCD. On the other hand there exist ample experimental support for the quark structure of hadrons. This is the origin of phenomenological models which are proposed to understand the properties of hadrons and quark dynamics at the hadronic scale. To first approximation, the confining part of the interaction is believed to provide the zeroth-order quark dynamics inside the meson through the quark Lagrangian density

\begin{equation}\label{eq:b}
{\cal L}^0_q (x)= \bar{\psi}_q(x) \left[\frac{i}{2} \gamma^{\mu} \overrightarrow{\partial_\mu} - V(r) - m_q \right] \psi_q(x).
\end{equation}
In this context for the present study, we assume that the constituent quark - antiquark inside a meson is independently confined by an average potential of the form \cite{Barik2000,Manan2014}
\begin{equation}\label{eq:a}
V(r)= \frac{1}{2} (1+\gamma_0) (\lambda r^{0.1}+V_0)
\end{equation}
In the stationary case, the spatial part of the quark wave functions $\psi(\vec{r})$ satisfies the Dirac equation given by
\begin{equation}\label{eq:c}
[\gamma^0 E_q - \vec{\gamma}. \vec{P} - m_q - V (r)]\psi_q (\vec{r}) = 0.
\end{equation}

The solution of Dirac equation can be written as two component (positive and negative energies in the zeroth order) form as
\begin{equation}\label{eq:d}
\psi_{nlj}(r) = \left(
    \begin{array}{c}
      \psi_{nlj}^{(+)} \\
      \psi_{nlj}^{(-)}
    \end{array}
  \right)
\end{equation}
where
\begin{equation}\label{eq:e}
\psi_{nlj}^{(+)}(\vec{r}) = N_{nlj} \left(
    \begin{array}{c}
      i g(r)/r \\
      (\sigma.\hat{r}) f(r)/r
    \end{array}
  \right) {\cal{Y}}_{ljm}(\hat{r})
\end{equation}
\begin{equation}\label{eq:f}
\psi_{nlj}^{(-)}(\vec{r}) = N_{nlj} \left(
    \begin{array}{c}
      i (\sigma.\hat{r}) f(r)/r \\
        g(r)/r
    \end{array}
  \right) (-1)^{j+m_j-l} {\cal{Y}}_{ljm}(\hat{r})
\end{equation}\\
and $N_{nlj}$ is the overall normalization constant. The normalized spin angular part is expressed as
\begin{equation}\label{eq:g}
{\cal{Y}}_{ljm}(\hat{r}) = \sum_{m_l, m_s}\langle l, m_l, \frac{1}{2}, m_s| j, m_j \rangle Y^{m_l}_l \chi^{m_s}_{\frac{1}{2}}
\end{equation}
Here the spinor $\chi_{\frac{1}{2}{m_s}}$ are eigenfunctions of the spin operators,
\begin{equation}\label{eq:h}
\chi_{\frac{1}{2} \frac{1}{2}} = \left(
    \begin{array}{c}
      1 \\
      0
    \end{array}
  \right) \ \ \ , \ \ \ \ \chi_{\frac{1}{2} -\frac{1}{2}} = \left(
    \begin{array}{c}
      0 \\
      1
    \end{array}
  \right)
\end{equation}
The reduced radial part $g(r)$ of the upper component and $f(r)$ of the lower component of Dirac spinor $\psi_{nlj}(r)$ are the solutions of the equations given by\\
\begin{equation}\label{eq:i1}
\frac{d^2 g(r)}{dr^2}+\left[(E_{D}+ m_q) [E_{D} - m_q - V(r)] - \frac{\kappa (\kappa + 1)}{r^2}\right] g(r) = 0
\end{equation}
and
\begin{equation}\label{eq:i2}
\frac{d^2 f(r)}{dr^2}+\left[(E_{D}+ m_q) [E_{D} - m_q - V(r)] - \frac{\kappa (\kappa - 1)}{r^2}\right] f(r) = 0
\end{equation}

\begin{table*}
\begin{center}
\caption{The fitted model parameters for the $D$ systems}\label{parameter}
\begin{tabular}{|c|c|}
\hline
\textbf{System Parameters } & \textbf{$D$}  \\
\hline\hline
Quark mass (in GeV)& $m_{u/d} = $ 0.003 and $m_c$ = 1.27 	   \\

\hline
Potential strength ($\lambda$) & $2.2903 + B $ GeV$^{\nu+1}$ \\
\hline
$V_0$     &- 2.6711 GeV    \\
\hline
Centrifugal parameter (B)  & $(n*0.153)$\ GeV$^{-1}$  for $l = 0$   \\
                           & $((n+l)*0.1267)$\ GeV$^{-1}$ for $l \neq 0$  \\
\hline
$\sigma$ ($j-j$ coupling strength) & 0.0055 \ GeV$^3$ for $l = 0$ \\
                                   & 0.0946 \ GeV$^3$ for $l \neq 0$ \\
\hline
\end{tabular}
\end{center}
\end{table*}


It can be transformed into a convenient dimensionless form given as \cite{barik1982}
\begin{equation}\label{eq:j1}
\frac{d^2 g(\rho)}{d\rho^2}+\left[\epsilon- \rho^{0.1} - \frac{\kappa (\kappa+1)}{\rho^2}\right] g(\rho) = 0
\end{equation}
 and
\begin{equation}\label{eq:j2}
\frac{d^2 f(\rho)}{d\rho^2}+\left[\epsilon- \rho^{0.1} - \frac{\kappa (\kappa -1)}{\rho^2}\right] f(\rho) = 0
\end{equation}
where $\rho = (r / r_0)$ is a dimensionless variable with the arbitrary scale factor chosen conveniently as
\begin{equation}\label{eq:k}
r_0 = \left[(m_q + E_{D})\frac{\lambda}{2}\right]^{-\frac{10}{21}},
\end{equation}
and $\epsilon$ is a corresponding dimensionless energy eigenvalue defined as
\begin{equation}\label{eq:l}
\epsilon = (E_{D} - m_q - V_0) (m_q + E_{D})^{\frac{1}{21}} \left(\frac{2}{\lambda}\right)^{\frac{20}{21}}
\end{equation}
Here, it is suitable to define a quantum number $\kappa $ by
\begin{eqnarray}\label{eq:kappa}
\kappa = &\left\{\begin{matrix} -(\ell + 1)& = - \left(j+\frac{1}{2}\right) & \ \ for \ \ j= \ell+\frac{1}{2}\\
                                      \ell & = + \left(j+\frac{1}{2}\right) & \ \ for \ \ j= \ell-\frac{1}{2} \end{matrix}\right.
\end{eqnarray}
Equations (\ref{eq:j1}) and (\ref{eq:j2}) now can be solved numerically \cite{Bhavin2009JPG} for each choice of $\kappa$.


The solutions $g (\rho)$ and $f (\rho)$ are normalized to get
\begin{equation}
 \int_0^\infty (f_q^2(\rho) + g_q^2(\rho)) \ d \rho = 1.
\end{equation}


The wavefunction for a $D (c\bar q)$  meson now can be constructed using Eqn (\ref{eq:e}) and (\ref{eq:f}) and the corresponding mass of the quark-antiquark system can be written as
\begin{equation}
M_{Q \bar q} \ (n_1l_1j_1, n_2l_2j_2)  =  E_{D}^Q + E_{D}^{\bar{q}}
\end{equation}
where $E_D^{Q/\bar{q}}$ are obtained using Eqn. (\ref{eq:l}) and (\ref{eq:kappa}) which include the centrifugal repulsion of the centre of mass. For the spin triplet (vector) and spin singlet (pseudoscalar) state, the choices of ($j_1$, $j_2$) are $\left(\left(l_1 + \frac{1}{2}\right), \left(l_2 + \frac{1}{2}\right)\right)$ and $\left(\left(l_{1, 2} + \frac{1}{2}\right), \left(l_{2, 1} - \frac{1}{2}\right)\right)$ respectively. The previous work of independent quark model within the Dirac formalism by \cite{Barik2000,Manan2014} has been extended here by incorporating the spin-orbit and tensor interactions of the confined one gluon exchange potential (COGEP) \cite{PCV1992,Khadkikar1991}, in addition to the j-j coupling of the quark-antiquark. Finally, the mass of the specific $^{2 S+1}L_J$ states of $Q \bar q$ system is expressed as
\begin{equation}\label{mass}
M_{^{2 S+1}L_J} =  M_{Q \bar q} \ (n_1l_1j_1, n_2l_2j_2) +  \langle V_{Q \bar q}^{j_1j_2}\rangle  + \langle V_{Q \bar q}^{LS}\rangle + \langle V_{Q \bar q}^{T}\rangle
\end{equation}

The spin-spin  part is defined here as
\begin{equation}
\langle V^{j_1 j_2}_{Q \bar q} (r)\rangle = \frac{\sigma \ \langle j_1 j_2 J M |\hat{j_1}.\hat{j_2}| j_1 j_2 J M \rangle}{(E_Q + m_{Q})(E_{\bar{q}} + m_{\bar{q}})}
\end{equation}
where $\sigma$ is the $j-j$ coupling constant. The expectation value of $\langle j_1 j_2 J M |\hat{j_1}.\hat{j_2}| j_1 j_2 J M \rangle$ contains the ($j_1.j_2$) coupling and the square of Clebsch-Gordan coefficients. The tensor and spin-orbit parts of confined one-gluon exchange potential (COGEP) \cite{PCV1992,Khadkikar1991} are given as
\begin{equation}\label{Vt}
V^{T}_{Q \bar q} (r) = - \frac{\alpha_s}{4} \frac{N_Q^2 N_{\bar q}^2}{(E_Q + m_{Q})(E_{\bar{q}} + m_{\bar{q}})} \otimes \ \lambda_Q . \lambda_{\bar q} \left( \left( \frac{D''_1 (r)}{3}- \frac{D'_1 (r)}{3 \ r} \right) S_{Q \bar q}\right)
\end{equation}
where $S_{Q \bar q} = \left[ 3 (\sigma_Q. {\hat{r}})(\sigma_{\bar q}. {\hat{r}})- \sigma_Q . \sigma_{\bar q}\right]$ and ${\hat{r}} = {\hat{r}}_Q - {\hat{r}}_{\bar q}$ is the unit vector in the direction of $\vec{r}$ and
\begin{eqnarray}\label{Vls}
V^{LS}_{Q \bar q} (r) &=& \frac{\alpha_s}{4} \frac{N_Q^2 N_{\bar q}^2}{(E_Q + m_{Q})(E_{\bar{q}} + m_{\bar{q}})}  \frac{\lambda_Q . \lambda_{\bar q}}{2 \ r} \\ &&\otimes \left[ \left[ \vec{r} \times (\hat{p_Q}-\hat{p_q}). (\sigma_Q + \sigma_q)\right]\left( {D'_0 (r)}+ 2 {D'_1 (r)} \right) \right. \nonumber\\ &&  \left. + \left[ \vec{r} \times (\hat{p_Q}+\hat{p_q}). (\sigma_i - \sigma_j)\right]\left( {D'_0 (r)}-  {D'_1 (r)} \right) \right]  \nonumber
\end{eqnarray}
where $\alpha_s$ is the strong coupling constant and it is computed as
 \begin{equation}
 \alpha_s = \frac{4 \pi}{(11-\frac{2}{3}\  n_\emph{f})\log\left(\frac{E^2_Q}{\Lambda^2_{QCD}}\right)}
 \end{equation}
 with $n_\emph{f}$ = 3 and $\Lambda_{QCD}$ = 0.150 GeV. In Eqs. (\ref{Vls}) the spin-orbit term has been split into symmetric $(\sigma_Q + \sigma_q)$ and anti-symmetric $(\sigma_Q - \sigma_q)$ terms.

We have adopted the same parametric form of the confined gluon propagators which are given by \cite{PCV1992,Khadkikar1991}
\begin{equation}
D_0 (r) = \left( \frac{\alpha_1}{r}+\alpha_2 \right) \exp(-r^2 c_0^2/2)
\end{equation}
and
\begin{equation}
D_1 (r) =  \frac{\gamma}{r} \exp(-r^2 c_1^2/2)
\end{equation}
with $\alpha_1$ = 0.036, $\alpha_2$ = 0.056, $c_0$ = 0.1017 GeV, $c_1$ = 0.1522 GeV, $\gamma$ = 0.0139 as in our earlier study \cite{Manan2014}. Other optimized model parameters employed in the present study are listed in Table \ref{parameter}. The current charm quark mass of 1.27 GeV is taken from the PDG (Particle data group)\cite{PDG2012}. In the case of $l \neq 0$ orbitally excited states, we find a small variations in the choice of $V_0$ for the $l = 0$ states due to the centrifugal repulsion from the center of mass of the bound system which is proportional to ($n+l$). This centrifugal repulsion thus incorporates the centre-of-mass correction.

\begin{figure}[tbp]
\centering
\includegraphics[width=0.90\textwidth]{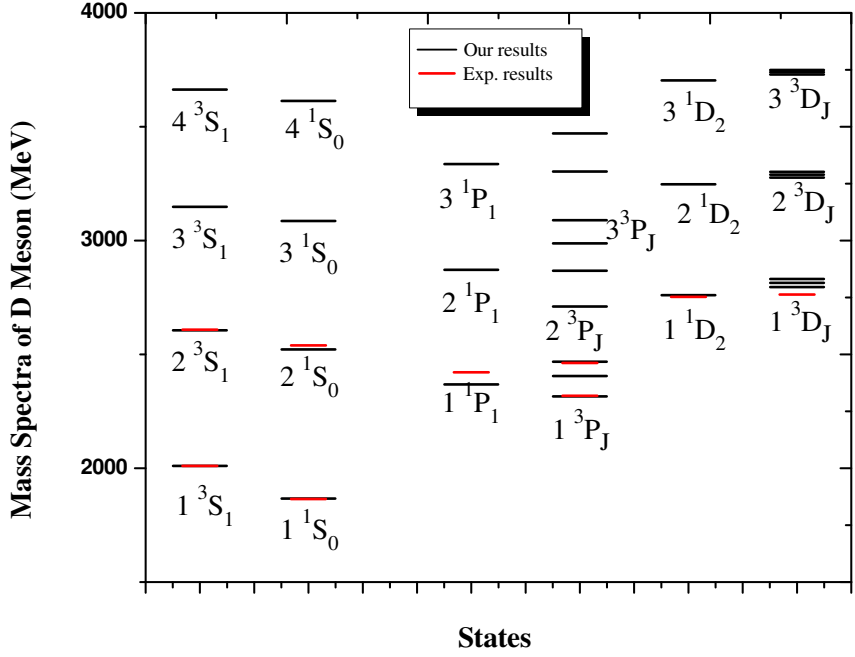}
\caption{\label{mass spectra}$D$ meson spectra.}
\end{figure}

The computed S-wave masses and other P-wave and D-wave masses of $D$ meson states are listed in Table \ref{tab1} and Table \ref{tab2} respectively. A statistical analysis of the sensitivity of the model parameters (i.e. potential strength ($\lambda$) and $j-j$ coupling strength $\sigma$ in the present case) shows about $0.76 \%$ variations in the binding energy with $5\%$ changes in the parameters $\lambda$ and $\sigma$. Fig.(\ref{mass spectra}) shows the energy level diagram of $D$ meson spectra along with available experimental results.\\

\begin{table}[tbp]
\begin{center}
\tabcolsep 1.5pt
 \small
\caption{S-wave $D$ ($c\bar{s}$) spectrum (in MeV).} \label{tab1}
\begin{tabular}{|c|c|c|c|c|c|c|c|c|c|c|c|c|c|c|c|}
\hline
 &    &  &       &&    & \multicolumn{2}{c|} {Experiment} &&& & &&&\\
\cline{7-8}
nL & $J^P$ & State &$M_{Q \bar q}$& $\langle V_{Q \bar q}^{j_1j_2}\rangle$ &Present  &  Meson  & Mass\cite{PDG2012} & \cite{Badalian2011}$^a$ & \cite{Ebert2010}$^b$ & \cite{Naynesh2013}$^c$ & \cite{De2011}$^d$& \cite{Yuan2013}$^e$ & \cite{Moir2013}$^f$ & QSR$^g$ \\
\hline
1S & $1^-$  &$1{^3S_1}$& 2009.54 & 0.99 &2010.53 & $D^*$   &  2010.28$\pm$0.13 &       & 2010 & 2018 & 2010& 2038 & 2013 & 2000$\pm$20 \cite{Hayashigaki2004}\\
   & $0^-$  &$1{^1S_0}$& 1869.57 &-2.58 &1867.00 & $D$     &  1864.86$\pm$0.13 &       & 1871 & 1865 & 1867& 1874 & 1890 & 1900$\pm$30 \cite{Hayashigaki2004}\\
\hline
2S & $1^-$  &$2{^3S_1}$& 2605.29 &0.57  &2605.86 &  $D^*$(2600) &  2608.7$\pm$2.4$\pm$2.5 \cite{Pdelamo2010} & 2639 & 2632 & 2639 & 2636 & 2645 & 2708 & 2612$\pm$6 \cite{Gelhausen2014}\\
   & $0^-$  &$2{^1S_0}$& 2523.05 &-1.33 &2521.72 &  $D$(2550)   &  2539.4$\pm$4.5$\pm$6.8 \cite{Pdelamo2010}  & 2567 & 2581 & 2598 & 2555 & 2583 & 2642 & 2539$\pm$8 \cite{Gelhausen2014} \\
\hline
3S & $1^-$  &$3{^3S_1}$& 3147.50 &0.39  &3147.89 &        &        & 3125 & 3096 & 3110 &   & 3111 & 3103 & \\
   & $0^-$  &$3{^1S_0}$& 3087.21 &-0.90 &3086.31 &        &        & 3065 & 3062 & 3087 &    & 3068 & 3064 & \\
\hline
4S & $1^-$  &$4{^3S_1}$& 3662.99 &0.29  &3663.28 &        &        &       & 3482 & 3514  &    &  & 3395 & \\
   & $0^-$  &$4{^1S_0}$& 3614.22 &-0.66 &3613.56 &        &        &       & 3452 & 3498 &      &  & 3299 & \\
\hline

\end{tabular}

\end{center}
\begin{flushleft}
    $^a$ Semi-relativistic model\\ $^b$ Quasi potential Approach \\ $^c$ Relativistic quark-antiquark potential (Coulomb plus power)
model\\ $^d$ Non-relativistic constituent quark model \\ $^e$ Relativistic quark model \\ $^f$ Lattice QCD [LQCD] \\ $^g$ QCD Sum Rule [QSR]
\end{flushleft}

\end{table}

\begin{table}[tbp]
\begin{center}
\tabcolsep 1.0pt
 \small
\caption{P-wave and D-wave $D$ ($c\bar{u}$ or $c\bar{d}$) spectrum (in MeV).} \label{tab2}
\begin{tabular}{|c|c|c|c|c|c|c|c|c|c|c|c|c|c|c|c|c|}
\hline
 &  &   &   &&&&           & \multicolumn{2}{c|}{Experiment}& &&& && &\\
\cline{9-10}
nL & $J^P$ & State &$M_{Q \bar q}$ &$\langle V_{Q \bar q}^{j_1j_2}\rangle$ &$ \langle V^{T}\rangle$& $\langle V^{LS}\rangle$& Present  &  Meson  & Mass \cite{PDG2012} & \cite{Badalian2011} & \cite{Ebert2010}& \cite{Naynesh2013} & \cite{De2011} & \cite{Yuan2013} & \cite{Moir2013} & \cite{Hayashigaki2004}\\
\hline
1P & $2^+$  &$1{^3P_2}$& 2411.01 &8.60  &$-$3.46  &52.07  &2468.22 & $D_{2}$(2460)   &  2462.6$\pm$0.7   &    & 2460 & 2473 & 2466 & 2501 & 2510 &\\
   & $1^+$  &$1{^3P_1}$& 2411.01 &28.68  &17.32  &$-$52.07 &2404.94 & $D_{1}$(2430)   &  2427$\pm$26$\pm$25 &    & 2469 & 2454 & 2417 & 2465 & 2478 & 2380$\pm$50\\
   & $0^+$  &$1{^3P_0}$& 2411.01 &43.02 &$-$34.65 &$-$104.14 &2315.24 & $D_{0}$(2400)   &  2318$\pm$29   &     & 2406 & 2352 & 2252 & 2398 & 2342 & 2450$\pm$30\\
   & $1^+$  &$1{^1P_1}$& 2312.60 &55.34  &0           & 0    &2367.94 &  $D_{1}$(2420)   &  2421.3$\pm$0.6  &     & 2426 & 2434 & 2402 & 2457 & 2446 &\\
\hline
2P & $2^+$   &$2{^3P_2}$& 2903.96 &5.89  &$-$5.57 &83.73   &2988.02 &                  &                      & 2965 & 3012 & 2971 & 2971 & 2957 & 3084 &\\
   & $1^+$   &$2{^3P_1}$& 2903.96 &19.65  &27.83 &$-$83.73  &2867.72 &                 &                      & 2960 & 3021 & 2951 & 2926 & 2952 & 3055 &\\
   & $0^+$   &$2{^3P_0}$& 2903.96 &29.47 &$-$55.67 &$-$167.46 &2710.31 &                 &                       & 2880 & 2919 & 2868 & 2752 & 2932 & 2996 &\\
   & $1^+$   &$2{^1P_1}$& 2835.21 & 36.31  &  0   &  0     &2871.51 &                 &                       & 2940 & 2932 & 2940 & 2886 & 2933 & 3051 &\\
\hline
3P & $2^+$  &$3{^3P_2}$& 3362.89 &4.43 &$-$7.37 &110.91  &3470.86 &                  &         &     & 3407 &     &    & & 3417 &\\
   & $1^+$  &$3{^3P_1}$& 3362.89 &14.76 &36.85 &$-$110.91 &3303.59 &                  &         &     & 3461 &     & &   & 3408 &\\
   & $0^+$  &$3{^3P_0}$& 3362.89 &22.14 &$-$73.71&$-$221.83 &3089.49 &              &         &     & 3346 &    &     & & 3351 &\\
   & $1^+$  &$3{^1P_1}$& 3309.13 &26.81  &  0  & 0       &3335.94 &                  &         &     & 3365 &     &    & & 3338 &\\
\hline
1D & $3^-$  &$1{^3D_3}$& 2839.42 &$-$8.51 &$-$0.02 &0.46 & 2831.34 &                 &         & 2840 & 2971 & 2834& 2811 & 2833 & 2870 &\\
   & $2^-$  &$1{^3D_2}$& 2839.42 &$-$25.30 & 0.08 &$-$0.23& 2813.97 &                  &                   & 2885 & 2961 & 2816 & 2788& 2834 & 2868 &\\
   & $1^-$  &$1{^3D_1}$& 2839.42 &$-$42.91&$-$0.08 &$-$0.69& 2795.74 &               &                   & 2870 & 2913 & 2873 & 2804& 2816 & 2850 &\\
   & $2^-$  &$1{^1D_2}$& 2761.19 &$-$1.04  & 0    & 0   & 2760.15 &     $D$(2750)   &  2752.4$\pm$1.7 & 2828 & 2931 & 2896 & 2849& 2827 & 2866 & \\
    &  &   &   &&&&           & &                                   $\pm$2.7 \cite{Pdelamo2010}                                             & &&& && &\\
\hline
2D & $3^-$  &$2{^3D_3}$& 3307.69 &$-$6.04 &$-$0.02 &0.45 & 3302.08 &                  &                   & 3285 & 3469 & 3263 & 3240 & 3226 & 3479 &\\
   & $2^-$  &$2{^3D_2}$& 3307.69 &$-$17.94 & 0.09 &$-$0.22& 3289.61 &                 &                   &      & 3456 & 3248 & 3217 & 3235 & 3426 &\\
   & $1^-$  &$2{^3D_1}$& 3307.69 &$-$30.42 &$-$0.09 &$-$0.68& 3276.51 &               &                   & 3290 & 3383 & 3292 & 3217 & 3231 & 3194 &\\
   & $2^-$  &$2{^1D_2}$& 3247.65 &$-$0.72  & 0    & 0   & 3246.93 &                   &                   &      & 3403 & 3312 & 3260 & 3225 & 3401 &\\
\hline
3D & $3^-$  &$3{^3D_3}$& 3753.22 &$-$4.58 &$-$0.03&0.52 & 3749.14 &                   &                   &      &      &    &   &  & &\\
   & $2^-$  &$3{^3D_2}$& 3753.22 &$-$13.60 & 0.10&$-$0.26& 3739.46 &                  &                   &      &      &     &  &  & &\\
   & $1^-$  &$3{^3D_1}$& 3753.22 &$-$23.06 &$-$0.10&$-$0.79& 3729.27 &                &                   &     &      &     &  &  & &\\
   & $2^-$  &$3{^1D_2}$& 3753.22 &$-$0.54  & 0   & 0   & 3703.34 &                    &                   &      &      &     &   &  & &\\
   \hline
\end{tabular}
\end{center}
\end{table}

\section{ Magnetic (M1) Transitions of Open Charm Meson}
Spectroscopic studies led us to compute the decay widths of energetically allowed radiative transitions of the type, $A \rightarrow B + \gamma $  among several vector and pseudoscalar states of $D$ meson. The magnetic transition correspond to spin flip and hence the vector meson decay to pseudoscalar $V\rightarrow P\gamma$ represents a typical M1 transition. Such transitions are experimentally important to the identification of newly observed states. Assuming that such transitions are single vertex processes governed mainly by photon emission from independently confined quark and antiquark inside the meson, the S-matrix elements in the rest frame of the initial meson is written in the form
\begin{equation}\label{eq:q}
S_{BA} = \left<B\gamma \left|-ie\int d^4 x \ T \left[\sum_q e_q \bar{\psi_q} (x) \gamma^\mu \psi_q (x) A_\mu (x) \right] \right|A \right>.
\end{equation}
The common choice of the photon field $A_\mu (x)$ is made here in Coulomb-gauge with $\epsilon (k, \lambda)$ as the polarization vector of the emitted photon having energy momentum $(k_0 = |\textbf{k}|,\textbf{k})$ in the rest frame of A. The quark field operators find a possible expansions in terms of the complete set of positive and negative energy solutions given by Eqs. (\ref{eq:e}) and (\ref{eq:f}) in the form
\begin{eqnarray}\label{eq:r}
\Psi_q (x) = \sum_\zeta &\left[b_{q\zeta} \ \psi_{q\zeta}^{(+)}(r) \ \exp(-iE_{q\zeta}t)  \right. \nonumber \\
  & \left. +\  b_{q\zeta}^{\dag} \ \psi_{q\zeta}^{(-)}(r) \ \exp(iE_{q\zeta}t)\right]
\end{eqnarray}
where the subscript q stands for the quark flavor and $\zeta$ represents the set of Dirac quantum numbers. Here $b_{q\zeta}$ and $b_{q\zeta}^{\dag}$ are the quark annihilation and the antiquark creation operators corresponding to the eigenmodes $\zeta$. After some standard calculations (the details of calculations can be found in Refs. \cite{Barik1992,Barik1993} and \cite{Jena1999}), the S-matrix elements can be expressed as
\begin{eqnarray}\label{eq:s}
S_{BA} &=& i \sqrt{\left(\frac{\alpha}{k}\right)}\  \delta (E_B +k-E_A) \sum_{q,m,m'} \left<B \left| \   \right. \right.  \\
  && \left. \left.  \left[ J^q_{m' m} (k, \lambda) b_{qm'}^\dagger b_{qm} - \  \tilde{J}^{\tilde{q}}_{m m'} (k, \lambda) \tilde{b}_{qm'}^\dagger \tilde{b}_{qm} \right] \right| A \right> \nonumber
\end{eqnarray}
Here $ E_A$ = $M_A$, $ E_B$ = $\sqrt{k^2 + M^2_B}$ and (m, m$'$) are the possible spin quantum numbers of the confined quarks corresponding to the ground state of the mesons. We have
\begin{equation}\label{eq:t}
J^q_{m' m} (k, \lambda) = e_q \int d^3 r \exp (-i \vec k \cdot \vec r) [ \bar{\psi}_{q m'}^{\ (+)} (r) \vec\gamma  \cdot \vec\epsilon (k, \lambda) \psi_{qm}^{(+)} (r)]
\end{equation}

\begin{equation}\label{eq:u}
\tilde{J}^{\tilde{q}}_{m m'} (k, \lambda)= e_q \int d^3 r \exp (-i \vec k \cdot \vec r) [ \bar{\psi}_{q m}^{\ (-)} (r) \vec\gamma  \cdot \vec\epsilon (k, \lambda) \psi_{qm'}^{(-)}(r)].
\end{equation}
One can reduce the above equations to simple forms as
\begin{equation}\label{eq:v}
J^q_{m' m} (k, \lambda) = - i\  \mu_q(k)\ [\chi_m^\dagger (\vec \sigma \cdot \vec K) \chi_m ],
\end{equation}
and
\begin{equation}\label{eq:v2}
\tilde{J}^{\tilde{q}}_{m m'} (k, \lambda) = i \ \mu_q(k)\ [\tilde{\chi}_m^\dagger (\vec \sigma \cdot \vec K) \tilde{\chi}_m ]
\end{equation}
where $\vec K = \vec k \times \vec \epsilon (k, \lambda)$. Eqs. (\ref{eq:s}) further simplified to get
\begin{eqnarray}\label{eq:w}
S_{BA} &=& i \sqrt{\left(\frac{\alpha}{k}\right)}\  \delta (E_B +k-E_A) \nonumber \\ && \sum_{q,m,m'} \left<B \left| \mu_q (k) \left[\chi_{m'}^\dagger \vec \sigma \cdot\vec K \chi_m b_{qm'}^\dagger b_{qm}  \right. \right. \right. \nonumber \\ && \left. \left. \left. + \  \tilde{\chi}_m^\dagger \vec \sigma \cdot\vec K \tilde{\chi}_{m'} \tilde{b}_{qm'}^\dagger \tilde{b}_{qm} \right] \right| A \right>
\end{eqnarray}
where $\mu_q(k)$ is expressed as
\begin{equation}\label{eq:x}
\mu_q(k)= \frac{2 e_q}{k} \int_0^\infty j_1(kr)\ f_q (r)\ g_q (r)\ dr
\end{equation}
where $j_1(kr)$ is the spherical Bessel function and the energy of the outgoing photon in the case of a vector meson undergoing a radiative transition to its pseudoscalar state, for instance, $D^* \rightarrow D \gamma$ is given by
\begin{equation}\label{eq:y}
k = \frac{M_{D^*}^2 - M_{D}^2}{2 M_{D^*}}
\end{equation}
The relevant transition magnetic moment is expressed as
\begin{equation}\label{eq:ab}
\mu_{D^+ D^{*+}}(k) = \frac{1}{3}[2 \mu_c (k) -  \mu_d (k)],
\end{equation}
\begin{equation}\label{eq:abb}
\mu_{D^0 D^{*0}}(k) = \frac{2}{3}[2 \mu_c (k) +  \mu_u (k)],
\end{equation}
Now, the Magnetic (M1) transition width of $D^* \rightarrow D \gamma$ can be obtained as
\begin{equation}\label{eq:ae}
\Gamma_{D^{*+} \rightarrow D^+ \gamma} = \frac{4 \alpha}{3} k^3 |\mu_{D^+ D^{*+}} (k)|^2
\end{equation}

\begin{equation}\label{eq:aee}
\Gamma_{D^{*0} \rightarrow D^0 \gamma} = \frac{4 \alpha}{3} k^3 |\mu_{D^0 D^{*0}} (k)|^2
\end{equation}
The computed transition widths of low lying S-wave states are tabulated in Table \ref{tab9} and are compared with other model predictions.

\begin{figure}[tbp]
\centering
\includegraphics[width=0.60\textwidth]{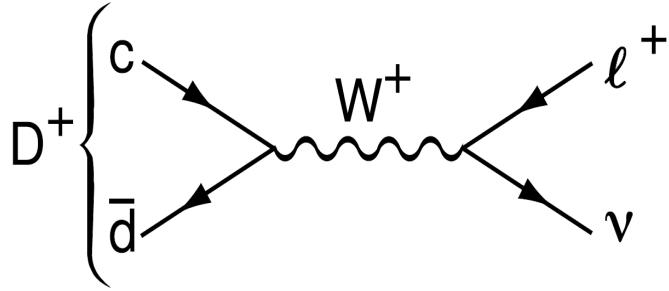}
\caption{Feynman diagram for leptonic decay $(M \rightarrow l \bar{\nu_l})$}\label{leptonic decay}
\end{figure}

\section{Decay constant of $D$ meson}
The decay constant of a meson is an important parameter in the study of leptonic or non-leptonic
weak decay processes. The decay constant ($f_p$) of pseudoscalar  state is obtained by parameterizing the matrix elements of weak current between the corresponding
meson and the vacuum as \cite{Quang}
\begin{equation}\label{eq:n}
\left<0|\bar{q}\gamma^{\mu} \gamma_5 c| P_\mu \right> = i f_p \ P^\mu
\end{equation}

\begin{table*}
\begin{center}
\caption{Comparison of Center of Mass in $D$ meson in MeV.}\label{tab12}
\begin{tabular}{|c|c|c|c|c|c|}
\hline
$M_{CW}$ 	      &	Present	&\cite{Naynesh2013}&\cite{Ebert2010} &	 Exp.	  \\
\hline				
$\overline{1S}$         & 1974.64	    & 1979.75   & 1975.25 &  1973.92  	\\
$\overline{2S}$         & 2584.82 	    & 2628.75   & 2619.25 &  2591.37    	 \\
$\overline{3S}$         & 3132.49 	    & 3104.25   & 3087.50 &          	 \\
$\overline{4S}$         & 3650.85 	    & 3510.00   & 3474.50 &           	 \\
$\overline{1^3P_J}$     & 2430.13 	    &  2453.22   & 2457.00 &   2434.66   	 \\
$\overline{1P}$         & 2414.58	    &  2448.42   & 2449.25 &  2431.22  \\	
$\overline{2^3P_J}$     & 2917.06 	    &  2952.88   & 3004.66 &      	 \\
$\overline{2P}$         & 2905.67 	    &  2949.66   & 2986.50 &         \\	
\hline
\end{tabular}
\end{center}
\end{table*}
\begin{table*}
\begin{center}
\caption{Mass splitting in $D$ meson in MeV.}\label{tab11}
\begin{tabular}{|c|c|c|c|c|c|c|c|}
\hline
Splitting 	       &	Present	&	\cite{Mohler2011}	& \cite{Naynesh2013} & \cite{Ebert2010} &	 Exp.	  \\
\hline		
$1^3S_1-1 ^1S_0$               &143.53	    &$130.8 \pm 3.2 \pm$ 1.8    & 153    &139  & $140.65 \pm 0.1$  	 \\
$2^3S_1-2 ^1S_0$               &84.14	    &                   & 41    &  51   &         	 \\
$3^3S_1-3 ^1S_0$               &61.58 	    &                  & 23    &34   &           	\\
$4^3S_1-4 ^1S_0$               &49.72 	    &                  & 16    &30   &            	 \\
$D_{0}$(2400)-$\overline{1S}$ &340.60      &$266.9 \pm 17.3 \pm$ 3.7  & 372.25   &430.75  & 347.0 $\pm$ 29\\
$D_{1}$(2420)-$\overline{1S}$ &393.30     &$399.1 \pm 13.5 \pm$ 5.6   & 454.25   &450.75  & 451.6 $\pm$ 0.6\\
$D_{1}$(2430)-$\overline{1S}$ &430.30     &$525.2 \pm 19.4 \pm$ 7.4   & 474.25   &493.75  & 456.0 $\pm$ 40\\
$D_{2}$(2460)-$\overline{1S}$ &493.58      &$577.1 \pm 20.3 \pm$ 8.1  & 493.25   &484.75  & 491.4 $\pm$ 1.0\\			

\hline

\end{tabular}
\end{center}
\end{table*}
It is possible to express the quark-antiquark eigenmodes in the ground state of the meson in terms of the corresponding momentum distribution amplitudes. Accordingly, eigenmodes, $\psi_A^{(+)}$ in the state of definite momentum p and spin projection $s'_p$ can be expressed as
\begin{equation}
\psi_A^{(+)} = \sum_{s'_p} \int d^3 p \ G_q(p,s'_p) \sqrt{\frac{m}{E_p}}\ U_q (p, s'_p) \exp (i\vec{p}\ .\ \vec{r})
\end{equation}
where $U_q (p, s'_p)$ is the usual free Dirac spinors.

In the relativistic quark model, the decay constant can be expressed
through the meson wave function $G_q (p)$ in the momentum
space \cite{Barik1993,HAKAN2000}
\begin{equation}\label{eq:p}
f_P = \left({\frac{3 | I_p |^2}{2 \pi^2 M_p\ J_p} }\right)^{\frac{1}{2}}
\end{equation}
Here $M_p$ is mass of the pseudoscalar meson and $I_p$ and $J_p$ are defined as
\begin{equation}
I_p = \int_0^\infty dp \ p^2 A (p) [G_{q1} (p) G^*_{q2} (-p)]^{\frac{1}{2}}
\end{equation}
\begin{equation}
J_p = \int_0^\infty dp \ p^2  [G_{q1} (p) G^*_{q2} (-p)]
\end{equation}
respectively. Where,
\begin{equation}
A (p) =  \frac{(E_{p1}+m_{q1})(E_{p2}+m_{q2})-p^2}{[E_{p1}\ E_{p2} (E_{p1}+m_{q1})(E_{p2}+m_{q2})]^{\frac{1}{2}}}
\end{equation}
and $E_{p_i} = \sqrt{{k_i}^2 + m_{q_i}^2}$.

The computed decay constants of $D$ meson from $1S$ to $4S$ states are tabulated in Table \ref{tab8}. Present result for $1S$ state is compared with experimental as well as other model predictions. There are no model predictions available for comparison of the decay constants of the $2S$ to 4S states.

\begin{table*}
\begin{center}
\caption{Magnetic (M1) transition of Open Charm Meson}\label{tab9}
\begin{tabular}{|c|c|c|c|c|c|c|c|c|}
\hline
 &  \multicolumn{2}{c|} {k (MeV)} & \multicolumn{6}{c|} {$\Gamma$ (keV)}\\
\cline{2-3} \cline{4-9}
Process & Present  &\cite{Naynesh2013} & $ Present$  & PDG \cite{PDG2012} &\cite{Naynesh2013}&\cite{Jena1998} & \cite{HAKAN2001} &  \cite{Ebert02} \\
\hline

(1S)$D^{*0} \rightarrow D^{0} \gamma$ &  138.38 & 147.00 & 1.2614  & $<$ 945 & 0.339 & 23.94  &   10.25  &  11.5    \\
(2S)$D^{*0} \rightarrow D^{0} \gamma$ &  82.84 & 41.00 & 0.0289 &            & 0.007   &      &         &      \\
(3S)$D^{*0} \rightarrow D^{0} \gamma$ &  60.99 & 23.00  & 0.0026 &           & 0.001  &      &          &     \\
(3S)$D^{*0} \rightarrow D^{0} \gamma$ &  49.46 & 16.00  & 0.0004 &           & 0.000  &      &           &    \\
\hline
(1S)$D^{*+} \rightarrow D^{+} \gamma$ &  138.38 & 147.00 & 0.0837  & $<$ 198 & 0.339 & 0.94  &   1.36   &  1.04   \\
(2S)$D^{*+} \rightarrow D^{+} \gamma$ &  82.84 & 41.00 & 0.0020 &            & 0.007   &      &         &      \\
(3S)$D^{*+} \rightarrow D^{+} \gamma$ &  60.99 & 23.00  & 0.0002 &           & 0.001  &      &          &     \\
(3S)$D^{*+} \rightarrow D^{+} \gamma$ &  49.46 & 16.00  & 0.0000 &           & 0.000  &      &          &    \\
\hline

\end{tabular}
\end{center}
\end{table*}

\begin{table*}
\begin{center}
\caption{Pseudoscalar  decay constant ($f_{P}$) of $D$ systems (in MeV).}\label{tab8}
\begin{tabular}{|c|c|c|c|c|c|c|c|}
\hline
 &  \multicolumn{4}{c|} {$f_{P}$}    \\
\cline{2-5}
 & 1S &  2S  & 3S &  4S   \\
\hline
Present                             & 202.57		    & 292.14       &351.066&  392.49 \\
PDG \cite{PDG2012}                  &	206.7 $\pm$ 8.9	&	           &      &         \\
$[CPP_{\nu}]$ \cite{Bhavin2010}      &	154 		&	           &      &     \\
$[QCDSR]$ \cite{Narison2013}         &	204 $\pm$ 6		&	           &      &     \\
$[RPM]$ \cite{Mao2012}               & 208 $\pm$ 21     &	           &      &          \\
$[QCDSR]$ \cite{Lucha2011}           &	206.2 $\pm$ 7.3 &	           &      &            \\
$[LQCD]$ \cite{Blossier2009}         & 197 $\pm$ 9	    &	           &      &          \\
$[LQCD]$ \cite{Bazavov2012}          & 218.9 $\pm$ 11.3 &	           &      &           \\
$[LFQM]$ \cite{Hwang2010}            & 206.0 $\pm$ 8.9  &	           &      &         \\
$[QCDSR]$ \cite{Wang2013}            & 208 $\pm$ 11     &	           &      &            \\
$[RBSM]$ \cite{Wang}                 & 229 $\pm$ 43     &	           &      &         \\
$[LQCD]$ \cite{Follana}              & 207 $\pm$ 11     &	           &      &         \\
$[LQCD]$ \cite{Heechang}             & 208 $\pm$ 3     &	           &      &         \\
\hline

\end{tabular}
\begin{center}
$[CPP_{\nu}]$- Coloumb plus power potential Model \\
$[QCDSR]$- QCD sum rule.\\
$[RPM]$- Relativistic potential Model.\\
$[LQCD]$- Lattice QCD.\\
$[LFQM]$- Light front quark model.\\
$[RBSM]$- Relativistic Bethe-Salpeter Method.\\
\end{center}
\end{center}
\end{table*}

\section{Leptonic Decay of $D$ Meson}
Charged mesons produced from a quark and anti-quark can decay to a charged lepton pair when
these objects annihilate via a virtual $W^\pm$ boson as given in Fig.(\ref{leptonic decay}). Though the leptonic decays of open flavour mesons belong to rare decay \cite{Hikasa1992,Rosner2008}, they have clear experimental signatures due to the presence of highly energetic lepton in the final state. And such decays are very clean due to the absence of hadrons in the final state \cite{Villa2007}. The leptonic width of $D$ meson is computed using the relation given by \cite{PDG2012}
\begin{eqnarray}\label{eq:af}
\Gamma(D \rightarrow l^+\nu_l)=\frac{G_F^2}{8\pi} f^2_{D} |V_{cd}|^2 m_l^2 \left(1-\frac{m_l^2}{M^2_{D}}\right)^2 M_{D}\ \ \ \ \
\end{eqnarray}
in complete analogy to $\pi^+ \rightarrow l^+ \nu$. These transitions are helicity suppressed; i.e., the amplitude is proportional to $m_l$, the mass of the lepton $l$. The leptonic widths of $D$ ($1^1S_0$) meson are obtained from Eqs.(\ref{eq:af}) where the predicted values of the pseudoscalar decay constant $f_{D}$ along with the masses of $M_{D}$ and the PDG value for $V_{cd}$ = 0.230 are used. The leptonic widths for separate lepton channel are computed for the choices of $m_{l=\tau, \mu, e}$. The branching ratio of these leptonic widths are then obtained as
\begin{equation}
BR = \Gamma (D \rightarrow l^+ \nu_l)\times \tau
\end{equation}
where $\tau$ is the experimental lifetime of the  respective $D$ meson state. The computed leptonic widths are tabulated in Table \ref{tab10} along with other model predictions as well as with the available experimental values. Our results are found to be in accordance with the reported experimental values.

\section{Hadronic Decays of $D$ Meson}


Study of flavour changing decays of heavy flavour quarks are useful for determining the parameters of the Standard Model and for testing phenomenological models which include strong effects. The interpretation of the hadronic decays of $c-$meson within a hadronic state is complicated by the effects of strong interaction and by its interplay with the weak interaction. The hadronic decays of heavy mesons can be understood in this model and we assume that Cabibbo favored hadronic decays proceed via the basic process, ($c\rightarrow q+ u +\bar{d}; q \in s,d$) and the decay widths are given by \cite{Quang}
\begin{eqnarray}\label{Hadronic1}
\Gamma (D^0\rightarrow K^- \pi^+) &=& C_f \frac{G_F^2 \ |V_{cs}|^2 |V_{ud}|^2 f_\pi^2 }{32 \ \pi \ M^3_{D_s}} \times \nonumber \\ &&[\lambda(M^2_{D},M^2_{K^-},M^2_{\pi})]^{\frac{3}{2}} |f^2_+(q^2)|
\end{eqnarray}
for $q=s$ and\\
\begin{eqnarray}\label{Hadronic2}
\Gamma(D^0\rightarrow K^+ \pi^-) &=& C_f \frac{G_F^2 \ |V_{cd}|^2 |V_{us}|^2  f_\pi^2 }{32 \ \pi \ M^3_{D_s}}  \times \nonumber \\ && [\lambda(M^2_{D},M^2_{K^+},M^2_{\pi})]^{\frac{3}{2}} |f^2_+(q^2)|
\end{eqnarray}
for $q=d$. Here, $C_f$ is the color factor and $(|V_{cs}|,|V_{cd}|, |V_{us}|)$ are the CKM matrices. $f_\pi$ is the decay constant of $\pi$ meson and its value is taken as 0.130 GeV. Here, $f_+(q^2)$ is the form factor and the factor $\lambda(M^2_{D},M^2_{K^+},M^2_{\pi})$ can be computed as
\begin{equation}
\lambda(x, y, z) = x^2 + y^2 + z^2 - xy - yz-zx
\end{equation}

The renormalized color factor without the interference effect due to QCD is given by $( C_A^2 + C_B^2)$. The coefficient $C_A$ and $C_B$ are further expressed as \cite{Quang}
\begin{equation}
C_A = \frac{1}{2}(C_+ + C_-)
\end{equation}
\begin{equation}
C_B = \frac{1}{2}(C_+ - C_-)
\end{equation}
where
\begin{equation}
C_{+} = 1 - \frac{\alpha_s}{\pi} \log \left(\frac{M_W}{m_c}\right)
\end{equation}
and
\begin{equation}
C_{-} = 1 + 2 \frac{\alpha_s}{\pi} \log \left(\frac{M_W}{m_c}\right)
\end{equation}
where $M_W$ is the mass of $W$ meson.\\
Consequently, the form factors  $f_{\pm}(q^2)$ correspond to the $D$ final state are related to the Isgur Wise function as \cite{Quang}
\begin{equation}
f_{\pm}(q^2) = \xi (\omega) \frac{M_{D}\pm M_{\phi}}{2 \sqrt{M_{D}M_{\phi}}}
\end{equation}
The Isgur Wise function, $\xi (\omega)$ can be evaluated according to the relation given by \cite{Olsson1995}
\begin{equation}
\xi (\omega) = \frac{2}{\omega-1}\left< j_0 \left( 2\ E_q \sqrt{\frac{\omega-1}{\omega+1}} \ r\right)\right>
\end{equation}
where $E_q$ is the binding energy of decaying meson and $\omega$ is given by,
\begin{equation}
\omega = \frac{M^2_{D}+ M^2_{(K^+, K^-)}-q^2}{2 M_{D}\ M_{(K^+, K^-)}}
\end{equation}

For a good approximation the form factor $f_-(q^2)$ do not contribute to the decay rate, so we have neglected here. The heavy flavour symmetry provides model-independent normalization of the weak form factors $f_{\pm}(q^2)$ either at $q=0$ or $q=q_{max}$ and we have applied $q=q_{max}$ in Eqs. (\ref{Hadronic1}) and (\ref{Hadronic2}) for hadronic decay. From the computed exclusive semileptonic and hadronic decay widths, the Branching ratios are obtained as
\begin{equation}\label{BR}
BR = \Gamma \times \tau
\end{equation}
here the lifetime ($\tau$) of $D$ $(\tau_{D^+} = 1.040 \ ps^{-1}$ and $\tau_{D^0} = 0.410 \ ps^{-1})$ is taken as the world average value reported by Particle Data Group (PDG-2012)\cite{PDG2012}.
The decay widths and their branching ratios  are listed in Table \ref{tab11} along with the known experimental and other theoretical predictions
for comparison.\\


\begin{table*}
\begin{center}
\tabcolsep 3.0pt
 \small
\caption{The leptonic decay width and leptonic Branching Ratio (BR) of $D$ meson.}
\label{tab10}
\begin{tabular}{|c|c|c|c|c|c|c|c|}
\hline
        &  \multicolumn{2}{c|} {$\Gamma(D^+ \rightarrow l \bar{\nu_l})$ (keV)}  &  \multicolumn{5}{c|} {BR }   \\
        \cline{2-3}\cline{4-8}
Process &  Present              & \cite{HAKAN2000} & Present & \cite{Naynesh2013} & \cite{HAKAN2000} & \cite{Bhavin2010} & Experiment \cite{PDG2012} \\
\hline
$D^+ \rightarrow \tau^+ {\nu_\tau}$  &6.157 $\times 10^{-10}$& 4.72 $\times 10^{-13}$ & 9.73 $\times 10^{-4}$&1.05 $\times 10^{-3}$ & 7.54 $\times 10^{-4}$ &  1.5 $\times 10^{-3}$      &   $<1.2\times 10^{-3}$ \\
$D^+ \rightarrow \mu^+ {\nu_\mu}$    &2.433 $\times 10^{-10}$& 1.79 $\times 10^{-13}$ & 3.84 $\times 10^{-4}$& 4.3 $\times
10^{-3}$ & 2.87 $\times 10^{-4}$ &  2.2 $\times 10^{-4}$          &  $3.82 \times 10^{-4}$ \\
$D^+ \rightarrow e^+ {\nu_e}$        &5.706 $\times 10^{-15}$&                    & 9.02 $\times 10^{-9}$& 1.00
$\times 10^{-8}$ &         &   0.5 $\times 10^{-8}$    & $<8.8 \times 10^{-6}$ \\
\hline
\end{tabular}
\end{center}
\end{table*}


\begin{table*}
\begin{center}
\tabcolsep 3.5pt
 \small
\caption{The Hadronic decay width and Branching Ratio (BR) of $D$ meson.}
\label{tab11}
\begin{tabular}{|c|c|c|c|c|c|c|}
\hline
  &  \multicolumn{1}{c|} {$\Gamma(D)$ (keV)}  &  \multicolumn{3}{c|} {BR}   \\
    \cline{2-2}\cline{3-5}
Process &  Present             & Present & \cite{Cheng2010}  & Experiment \cite{PDG2012}  \\
\hline
$D^0 \rightarrow K^- \ \pi^+ $      &6.153 $\times 10^{-14}$  & $3.835 \times 10^{-2}$  & $(3.91 \pm 0.17)\%$ & $(3.91 \pm 0.08)\% $ \cite{Mendez2010}\\
$D^0 \rightarrow {K}^+ \ \pi^-$     &1.716 $\times 10^{-16}$  & $1.069 \times 10^{-4} $ & ($1.12  \pm 0.05) \times 10^{-4} $ &$(1.48 \pm 0.07)\times 10^{-4}$ \cite{Mendez2010}\\

\hline
\end{tabular}
\end{center}
\end{table*}

\section{Mixing Parameters of $D - \bar{D}$ Oscillation }
 A different $D^0$ decay channel \cite{Aubert2007,Staric2007,Aaltonen2008,Aubert20091,Aubert20092} has been reported by three experimental groups as evidence of $D^0 - \bar{D}^0$
oscillation. We discuss here the mass oscillation of the neutral open charm meson and the integrated oscillation rate using our spectroscopic parameters deduced from the present study. In the standard model, the transitions $D^{0}-\bar{D}^{0}$ and $\bar{D}^{0}-{D}^{0}$  occur through the weak
interaction. The neutral $D$  meson mix with their
antiparticle leading to oscillations between the mass eigenstates
\cite{PDG2012}. In the following, we adopt the notation introduced
in \cite{PDG2012}, and assume $CPT$ conservation in our calculations. If CP symmetry
is violated, the oscillation rates for meson produced as $D^0$
and $\bar{D}^0$ can differ, further enriching the phenomenology. The study of CP violation
in $D^0$ oscillation may lead to an improved understanding of
possible dynamics beyond the standard model \cite{Blaylock1995,Petrov2006,Golowich2007}.\\

The time evolution of the neutral $D-$meson
doublet is described by a Schr$\ddot{o}$dinger equation with an
effective $2\times2$ Hamiltonian given by \cite{Quang,G.Buchalla2008}
\begin{equation}
i\frac{d}{dt}\left(
  \begin{array}{c}
    D^{0}(t) \\
   \bar{D}^{0}(t) \\
  \end{array}
\right)=\left(
           \begin{array}{cc}
             M- \frac{i}{2}\Gamma
           \end{array}
         \right) \left(
  \begin{array}{c}
    D^{0}(t) \\
   \bar{D}^{0}(t) \\
  \end{array}
  \right)
\end{equation}

where the \textbf{M} and \textbf{$\Gamma$} matrices are Hermitian, and are defined as
\begin{equation}
\left(
           \begin{array}{cc}
             M- \frac{i}{2}\Gamma
           \end{array}
         \right) =\left[\left(
           \begin{array}{cc}
             M^{q}_{11} & M^{q*}_{12} \\
             M^{q}_{12} & M^{q}_{11}\\
           \end{array}
         \right)
- \frac{i}{2}\left(
           \begin{array}{cc}
             \Gamma^{q}_{11} & \Gamma^{q*}_{12} \\
             \Gamma^{q}_{12} & \Gamma^{q}_{11}\\
           \end{array}
         \right) \right].
\end{equation}

CPT invariance imposes
\begin{equation}
M_{11} = M_{22} \equiv M , \Gamma_{11} = \Gamma_{22} \equiv \Gamma.
\end{equation}

The off-diagonal elements of these matrices describe the dispersive and absorptive parts of $D^0 - \bar{D}^0$
mixing \cite{Bigi1986}. The two eigenstates $D_1$ and $D_2$ of the
effective Hamiltonian matrix $ (M- \frac{i}{2}\Gamma)$ are given by
\begin{equation}
|D_1\rangle = \frac{1}{\sqrt{|p|^2+|q|^2}} (p|D^0\rangle+ q|\bar{D}^0\rangle),
\end{equation}
\begin{equation}
|D_2\rangle = \frac{1}{\sqrt{|p|^2+|q|^2}} (p|D^0\rangle - q|\bar{D}^0\rangle).
\end{equation}
The corresponding eigenvalues are
\begin{equation}\label{ld1}
\lambda_{D_1} \equiv   m_1 - \frac{i}{2}\Gamma_1 = \left( M- \frac{i}{2}\Gamma\right) + \frac{q}{p} \left( M_{12} - \frac{i}{2}\Gamma_{12} \right),
\end{equation}
\begin{equation}\label{ld2}
\lambda_{D_2} \equiv   m_2 - \frac{i}{2}\Gamma_2 = \left( M- \frac{i}{2}\Gamma\right) - \frac{q}{p} \left( M_{12} - \frac{i}{2}\Gamma_{12} \right),
\end{equation}
where $m_1(m_2)$ and $\Gamma_1(\Gamma_2)$ are the mass and width of $D_1 (D_2)$, respectively, and
\begin{equation}
\frac{q}{p} = \left( \frac{M_{12}^* - \frac{i}{2}\Gamma_{12}^*}{{M_{12} - \frac{i}{2}\Gamma_{12}}} \right)^{1/2}
\end{equation}
From Eqs. (\ref{ld1}) and (\ref{ld2}), one can get the differences in mass and width which are given as
\begin{equation}
\Delta m \equiv m_2 - m_1 = - 2 Re \left[\frac{q}{p} (M_{12} - \frac{i}{2}\Gamma_{12}) \right],
\end{equation}
\begin{equation}
\Delta \Gamma \equiv \Gamma_2 - \Gamma_1 = - 2 Im \left[\frac{q}{p} (M_{12} - \frac{i}{2}\Gamma_{12}) \right].
\end{equation}

\begin{figure}[tbp]
\centering
\includegraphics[width=.45\textwidth]{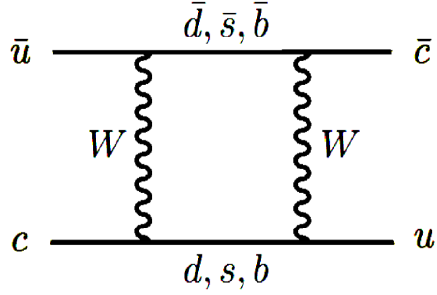}
\caption{$ D^0 - \bar{D}^0$ mixing }\label{mixing}
\end{figure}

The calculation of the dispersive and absorptive parts of the box
diagrams yields the following expressions for the off-diagonal
element of the mass and decay matrices; for example, if $ s/\bar{s}$ as the intermediate quark state then \cite{Buras84},
\begin{eqnarray} 
\hspace{-4.0ex} M_{12} \hspace{-1ex}\ &=&\ \hspace{-1ex} - \frac{
           G_F^2 m_W^2 \eta_{D^{0}} m_{D^{0}} B_{D^{0}} f_{D^{0}}^2}{12\pi^2}
            S_0(m_s^2/m_W^2) (V_{us}^* V_{cs}^{})^2 \,
 \\
\hspace{-4.0ex} \Gamma_{12} \hspace{-1ex}\ &=&\ \hspace{-1ex} \frac{
           G_F^2 m_c^2 \eta_{D^{0}}^{'} m_{D^{0}} B_{D^{0}} f_{D^{0}}^2}{8\pi}
\hspace{-0.5ex}
           \left[ (V_{us}^* V_{cs}^{})^2
           \right] \hspace{-0.5ex}\,
\end{eqnarray}
where $G_F$ is the Fermi constant, $m_W$ is the $W$ boson mass, $m_c$
is the mass of $c$ quark, $m_{D^{0}}$, $f_{D^{0}}$ and $B_{D^{0}}$ are the
$D^0$ mass, weak decay constant and bag parameter respectively. The
known function $S_0(x_q)$ can be approximated very well by
$0.784\,x_q^{0.76}$  \cite{Buras_Fleischer_HeavyFlavorsII} and
$V_{ij}$ are the elements of the CKM matrix \cite{CKM}. The
parameter $\eta_D^{0}$ and $\eta'_{D^{0}}$ correspond to the glunoic
corrections. The only non-negligible contributions to $M_{12}$ are
from box diagrams involving $s (\bar{s}), d (\bar{d}), b(\bar{b})$ intermediate quarks in Fig. (\ref{mixing}). The phases of $M_{12}$
and $\Gamma_{12}$ satisfy
\begin{equation} \label{eq:phase} 
\phi_{M} - \phi_{\Gamma} = \pi + {\cal O}\left(\frac{m^2_c}{m^2_b}
\right) \,,
\end{equation} 
implying that the mass eigenstates have mass and width differences
of opposite signs. This means that, like in the
$K{^0}\hbox{--}\overline K{^0}$ system, the heavy state is expected
to have a smaller decay width than that of the light state:
$\Gamma_{1} < \Gamma_{2}$. Hence, $\Delta \Gamma =
\Gamma_{2} -\Gamma_{1}$ is expected to be positive in the
Standard Model. Further, the quantity
\begin{equation}\label{eq:phase12} 
\left|\frac{\Gamma_{12}}{M_{12}}\right| \simeq \frac{3\pi}{2}
\frac{m^2_c}{m^2_W} \frac{1}{S_0(m_q^2/m_W^2)} \sim {\cal
O}\left(\frac{m^2_q}{m^2_t} \right)
\end{equation} 
is small, and a power expansion of $|q/p|^2$ yields
\begin{equation} 
\left|\frac{q}{p}\right|^2 = 1 +
\left|\frac{\Gamma_{12}}{M_{12}}\right| \sin(\phi_{M}-\phi_{\Gamma})
+ {\cal O}\left( \left|\frac{\Gamma_{12}}{M_{12}}\right|^2\right)
\,.
\end{equation} 
Therefore, considering both
Eqs.(\ref{eq:phase}) and (\ref{eq:phase12}), the $CP$-violating
parameter given by
\begin{equation} 
1 - \left|\frac{q}{p}\right|^2 \simeq {\rm
Im}\left(\frac{\Gamma_{12}}{M_{12}}\right)
\end{equation} 
is expected to be very small: $\sim {\cal O}(10^{-3})$ for the $D^0-
\bar D^0$ system. In the approximation of negligible $CP$
violation in mixing, the ratio $\Delta\Gamma/\Delta m$ is equal
to the small quantity $\left|\Gamma_{12}/M_{12}\right|$ of Eqs.
(\ref{eq:phase12}); it is hence independent of CKM matrix elements,
{\it i.e.}, the same for the $D^0- \bar D^0$ system.\\
\\Theoretically, the hadron lifetime ($\tau_{D^{0}}$) is related to $\Gamma_{11} (
\tau_{D^{0}}=1/ \Gamma_{11})$, while the observable $\Delta m$ and  $\Delta
\Gamma$ are related to $M_{12}$ and $\Gamma_{12}$ as \cite{PDG2012}
\begin{equation}
\Delta m =2 |M_{12}|
\end{equation}
and
\begin{equation}
\Delta\Gamma=2 |\Gamma_{12}|
\end{equation}

The gluonic correction can find from by different model like Wilson coefficient and evolution of Wilson coefficient from the new physics scale \cite{Golowich2007}. We have used values of gluonic correction ($\eta_{D^0}= 0.86$; $\eta_{D^0}^{'} = 0.21$) from \cite{Monika2007, Stephan1994}. The bag parameter $B_{D^{0}}= 1.34$ is taken from
the lattice result of \cite{Buras2003}, while the pseudoscalar mass
($M_{D^{0}}$) and the pseudoscalar decay constant ($f_{D}$) of
$D$ mesons are the values obtained from our present study  using relativistic independent quark
model using Martin like potential. The values of $m_{s}$ (0.1 GeV), $M_{W}$ (80.403
GeV) and the  CKM matrix elements $V_{cs} (1.006)$ and $V_{us}
(0.2252)$ are taken from the Particle Data Group \cite{PDG2012}. The resulting mass
oscillation parameter $\Delta m$ are tabulated in
Table-\ref{tab:LM08} with latest experimental
results.\\
\\The integrated oscillation rate ($\chi_q$) is the probability to
observe a $\bar D$ meson in a jet initiated by a $\bar c$ quark. As
the mass difference $\Delta m_{D}$ is a measure of the frequency of
the change from a $D^{0}$ into a $\bar D^{0}$ or vise versa.  This
change is reflected in either the time-dependent oscillations or in
the time-integrated rates corresponding to the di-lepton events
having the same sign. The time evolution of the neutral states from
the pure $|D{^0}_{phys}\rangle$ or $| \bar{D}{^0}_{phys}\rangle$ state at
$t=0$ is given by
\begin{eqnarray} \label{time_evol1} 
| D{^0_{phys}}(t)\rangle &=& g_+(t) \,| D{^0}\rangle
                     + \frac{q}{p} \, g_-(t) \,| \bar{D}{^0}\rangle \,,
\\
| \bar{D}{^0_{phys}}(t)\rangle &=& g_+(t) \,| \bar{D}{^0}\rangle
                     + \frac{p}{q} g_-(t) \,| D{^0}\rangle \,,
\end{eqnarray} 
which means that the flavor states remain unchanged ($g_+$) or
oscillate into each other ($g_-$) with time-dependent probabilities
proportional to
\begin{equation} \label{cosh1}
 g_{+}(t)\ = e^{ \frac{-\Gamma t}{2}}  e^{ {-i\ t\ m_{D^0}}}  \cos(t \Delta
m/ 2) ,
\end{equation} 
\begin{equation} \label{cosh2}
 g_{-}(t)\ = e^{ \frac{-\Gamma t}{2}}  e^{ {-i\ t\ m_{D^0}}}  \sin(t \Delta
m/ 2).
\end{equation} 
Starting at $t = 0$ with initially pure $D^0$, the probability for finding a $D^0 (\bar{D}^0)$ at time $t \neq 0$ is
 given by $\left|g_+ (t)\right|^2 (\left|g_- (t)\right|^2 )$. Taking $|q/p| = 1 $, one gets
\begin{equation} \label{cosh3}
 \left|g_\pm (t) \right|^2\ = \frac{1}{2} e^{ \frac{-\Gamma_D t}{2}}  [1 \pm   \cos(t \Delta
m)].
\end{equation}
Conversely, from an initially pure $\bar{D}^0$ at $t = 0 $, the probability for finding a $\bar{D}^0 (D^0)$ at time
 $t \neq 0$ is also given by  $\left|g_+ (t)\right|^2 (\left|g_- (t)\right|^2 )$.The oscillation of $D^0$ or $\bar{D}^0$
 as shown by Eqs. (\ref{cosh3}) give $ \Delta m $ directly. Integrating $\left|g_\pm (t) \right|^2$ from $t = 0$ to $ t = \infty$, we get
\begin{equation} \label{cosh4}
\int^\infty_0 \left|g_\pm (t) \right|^2 dt \ = \frac{1}{2} \left[ \frac{1}{\Gamma} \pm \frac{\Gamma}{\Gamma^2+(\Delta m)^2}\right]
\end{equation}
where $\Gamma = \Gamma_D = (\Gamma_{1} +\Gamma_{2})/2$.
The ratio
\begin{equation} \label{cosh5}
 r_o = \frac{D^0 \leftrightarrow \bar{D}^0}{D^0 \leftrightarrow {D}^0} = \frac{\int^\infty_0 \left|g_- (t) \right|^2 dt}{\int^\infty_0 \left|g_+ (t) \right|^2 dt} = \frac{x^2}{2 + x^2},
\end{equation}
\begin{eqnarray} \label{chi} 
~~~{\rm where}~~~ x_q = x =
\frac{\Delta m}{\Gamma}= \Delta m\ \tau_{D},  ~~~ y_q =
\frac{\Delta \Gamma }{2\Gamma}=\frac{\Delta \Gamma
\ \tau_{D}}{2}    \nonumber,
\end{eqnarray} 
\begin{equation}
 \chi_q = \frac{x_q^2+y_q^2}{2(x_q^2+1)},
\end{equation}
reflects the change of pure $D^0$ into a $\bar{D}^0$, or vice versa.\\
The time-integrated mixing rate relative to the time-integrated right-sign decay rate for
semileptonic decays \cite{PDG2012} is
\begin{equation}
 R_M = \int^\infty_0 r (t)  dt  = \int^\infty_0 \left|g_- (t) \right|^2  \left|\frac{q}{p} \right|^2 dt
\end{equation}
\begin{equation}
 R_M = \int^\infty_0 \frac{e^{-t}}{4} (x^2 + y^2) t^2 \left|\frac{q}{p} \right|^2  \simeq \frac{1}{2} (x^2 + y^2)
\end{equation}

In the Standard Model, CP violation in charm mixing is small and $|q/p|$ $\approx$ 1.

For the present estimation of these mixing parameteres $x_q$, $y_q$
and $\chi_q$, we employ our predicated $\Delta m$ values and
the experimental average lifetime of PDG \cite{PDG2012} of the
$D$-meson.

\begin{table*}
\begin{center}
\caption{Mixing Parameters $x_q$, $y_q$, $\chi_q$ and $R_M$ of  $D$ mesons}
\label{tab:LM08}
\begin{tabular}{|c|c|c|c|c|c|c|c|}
\hline
 & $\Delta M (GeV) $ &{$x_q$} & {$y_q$}& {$\chi_q$} & $R_M$\\

  \hline
    Present   & 8.255 $\times 10^{-15}$  &5.14 $\times 10^{-3}$ &   6.02$\times 10^{-3}$  &  3.13$\times 10^{-5}$ & 3.13 $\times 10^{-5}$   \\
    \cite{Zhang2007}  &   &(0.80 $\pm$ 0.29)$\%$    & (0.33 $\pm$ 0.24)$\%$   &   &   0.864 $ \pm$ 0.311 $\times 10^{-4}$    \\
    \cite{Bitenc2008}  & &     &         &     & 0.13$\pm$0.22$\pm$0.20 $\times 10^{-3}$       \\
     \cite{Aubert20072}  &&       &        &   &   0.04$^{+0.7}_{-0.6}$ $\times 10^{-3}$       \\
 \cite{Bitenc2005}  & &      &         &     & 0.02$\pm$0.47$\pm$0.14 $\times 10^{-3}$       \\

\hline
\end{tabular}
\end{center}
\end{table*}

\section{Results and Discussion}
We have studied here the mass spectra and decay properties of the $D$ meson in the framework of relativistic independent quark model. Our computed $D$ meson spectral states are in good agreement with the reported PDG values of known states.
The predicted masses of S-wave $D$ meson state $2\ ^3S_1$ (2605.86 MeV) and $2\ ^1S_0$ (2521.72 MeV) are in very good agreement with the respective experimental results of $2608\pm{2.4}\pm{2.5}$ MeV \cite{Pdelamo2010} and $2539.4\pm{4.5}\pm{6.8}$ MeV \cite{Pdelamo2010}  by BABAR Collaboration. The expected results of other S-wave excited states of $D$ meson are also in good agreement with other reported values  \cite{Badalian2011,Ebert2010,Naynesh2013,De2011}. The predicted P-wave $D$ meson states, $1^3P_2$ (2468.22 MeV), $1^3P_1$ (2404.94 MeV), $1^3P_0$ (2315.24 MeV) and $1^1P_1$ (2367.94 MeV) are in good agreement with experimental \cite{PDG2012} results of $2462.6 \pm 0.7 $ MeV, $2427 \pm 26 \pm 25$ MeV, $2318 \pm 29 $ MeV and $2421.3 \pm 0.6 $ MeV respectively. The $1 ^1D_2$ (2760.15) are very close with the experimental result of $2752.4\pm1.7\pm2.7$ MeV \cite{Pdelamo2010} and we predict its $J^P$ value to be $2^{-}$. We have also compared Lattice QCD and QCD sum rule results with our predicted results where the numerical values in Table \ref{tab2} for Lattice  QCD results are extracted from the energy level diagram available in \cite{Moir2013}. With reference to the available experimental masses of $D-$mesonic states, we observe that the LQCD predictions \cite{Moir2013} are off by standard deviation of $\pm$ 58.52 and that by QCD sum rule \cite{Gelhausen2014,Hayashigaki2004} predictions are off by $\pm$ 59.22, while predicted calculations show the standard deviation of $\pm$ 21.88.   \\

In the relativistic Dirac formalism, the spin degeneracy is primarily broken therefore to compare the spin average mass, we employ the relation as
\begin{equation}
M_{CW} = \frac{\sum_J (2 J +1) M_J}{\sum_J (2 J +1)}
\end{equation}
The spin average or the center of weight masses $M_{CW}$ are calculated from the known values of the different meson states and are compared with other model predictions \cite{Ebert2010} and \cite{Naynesh2013} in Table \ref{tab12}. The table also contains the different spin dependent contributions for the observed state.\\

The precise experimental measurements of the masses of $D$ meson states provided a real test for the choice of the hyperfine and the fine structure interactions adopted in the study of $D$ meson spectroscopy. Recent study of  $D$ meson mass splitting in lattice QCD [LQCD] \cite{Mohler2011} using 2 $\pm$ 1 flavor configurations generated with the Clover-Wilson fermion action by the PACS-CS collaboration \cite{Mohler2011} has been listed for comparison. Present results as seen in Table \ref{tab11} are in very good agreement with the respective experimental values over the lattice results \cite{Mohler2011}. In this Table, the present results on an average, are in agreement with the available experimental value within $12 \%$ variations, while the lattice QCD predictions \cite{Mohler2011} show $30 \%$ variations.\\

The magnetic transitions (M1) can probe the internal charge structure of hadrons, and therefore they will likely play an important role in determining the hadronic structures of $D$ meson. The present M1 transitions widths of $D$ meson states as listed in Table \ref{tab9} are in accordance with the model prediction of \cite{HAKAN2001} while the upper bound provided by PDG \cite{PDG2012} is very wide. We do not find any theoretical predictions for M1 transition width of excited states for comparison. Thus we only look forward to see future experimental support to our predictions. \\

The calculated pseudoscalar decay constant ($f_P$) of $D$ meson is listed in Table (\ref{tab8}) along with other model predictions as well as experimental results. The value of $f_{D} (1S)$ = 202.57 MeV obtained in our present study is in very good agreement with other theoretical predictions for $1S$ state.  The predicted $f_{D}$ for higher S-wave states are found to increase with energy. However, there are no experimental or theoretical values available for comparison.  Another important property of $D$ meson studied in the present case is the leptonic decay widths. The present branching ratios for $D \rightarrow \tau \bar{\nu_\tau}$ ($9.73 \times 10^{-4}$) and $D \rightarrow \mu \bar{\nu_\mu}$ ($3.846 \times 10^{-4}$) are in accordance with the experimental results $ (<1.2 \times 10^{-2})$ and  $(3.82 \times 10^{-4})$ respectively over other theoretical predictions vide Table \ref{tab10}. Large experimental uncertainty in the electron channel make it difficult for any reasonable conclusion.\\

The Cabibbo favoured hadronic branching ratio BR$(D^0\rightarrow K^- \pi^+)$ and BR $(D^0\rightarrow K^+ \pi^-)$ obtained respectively as $3.835 \%$ and  $1.069 \times 10^{-4} $ are in very good agreement with Experimental values \cite{Mendez2010} of $ 3.91 \pm 0.08\%$ and $(1.48\pm 0.07) \times 10^{-4}$ respectively.\\

We obtained the CP violation parameter in mixing $|q/p|$ (0.9996) in this case and $D^0$ and $\bar{D}^0$ decays shows no evidence for CP violation and provides the most stringent bounds on the mixing parameters. The mixing parameter $x_q$, $y_q$, and mixing rate $(R_M)$ are very good agreement with BaBar, Belle and other Collaboration as shown in Table \ref{tab:LM08}.  However, due to larger uncertainty in the experimental values make difficult for us to draw a continuous remark on this mixing parameter. Thus, the present study of the mixing parameters of neutral open charm meson is found to be one of the successful attempt to extract the effective quark-antiquark interaction in the case of heavy-light flavour mesons. Thus the present study is an attempt to indicate the importance of spectroscopic (strong interaction) parameters in the weak decay processes.\\

Finally we look forward to see future experimental support in favour of many of our predictions on the spectral states and decay properties of the open charm meson.

\acknowledgments

This work is part of Major Research Project No. F. 40-457/2011(SR) funded by UGC, India. One of the authors (Bhavin Patel) acknowledges the support through the Fast Track project funded by DST (SR/FTP/PS-52/2011).


\end{document}